\documentclass[prd, twocolumn, nofootinbib, superscriptaddress]{revtex4-2}

\usepackage{aas_macros}
\usepackage{graphicx}
\usepackage[caption=false]{subfig}
\usepackage{siunitx}
\usepackage{mathrsfs}
\usepackage{amsmath,amssymb}
\usepackage{bm}
\usepackage{braket}
\usepackage{listings}
\usepackage{cases}
\usepackage{here}
\usepackage{comment}
\usepackage{soul}
\usepackage{cancel}
\usepackage{cases}
\usepackage[utf8]{inputenc}
\usepackage{url}
\usepackage{longtable}
\usepackage[normalem]{ulem}
\usepackage{xspace}
\usepackage[colorlinks=true
,urlcolor=blue
,anchorcolor=blue
,citecolor=blue
,filecolor=blue
,linkcolor=blue
,menucolor=blue
,linktocpage=true
,pdfproducer=medialab
,pdfa=true
]{hyperref}
\usepackage{animate}
\usepackage{mathcomp}

\begin{document}


\title{Constraints on primordial black holes from the observed number of Icarus-like ultrahigh magnification events}

\author{Hiroki Kawai}
\email{hiroki.kawai@phys.s.u-tokyo.ac.jp}
\affiliation{Department of Physics, The University of Tokyo, Bunkyo, Tokyo 113-0033, Japan}
\affiliation{Center for Frontier Science, Chiba University, 1-33 Yayoicho, Inage, Chiba 263-8522, Japan}
\affiliation{INAF-Osservatorio di Astrofisica e Scienza dello Spazio di Bologna, Via Piero Gobetti 93/3, 40129 Bologna, Italy}

\author{Masamune Oguri}
\affiliation{Center for Frontier Science, Chiba University, 1-33 Yayoicho, Inage, Chiba 263-8522, Japan}
\affiliation{Department of Physics, Graduate School of Science, Chiba University, 1-33 Yayoicho, Inage, Chiba 263-8522, Japan}

\date{\today}

\begin{abstract}
Icarus is an individual star observed near the macro-critical curve of the MACS J1149 cluster, with the magnification factor estimated to be an order of thousands.
Since microlenses near the macro-critical curve influence the number of such high-magnification events, the observed occurrence of Icarus-like events is expected to provide a useful constraint on the properties of microlenses.
We first study the mass and mass fraction of microlenses consistent with the observed number of events assuming a single microlens component with a monochromatic mass function, finding that stars that contribute to the intracluster light (ICL) are consistent at the 95\% confidence level. 
We then consider the contribution of primordial black holes (PBHs), which are one of the alternatives to the standard cold dark matter, as microlenses in addition to ICL stars. 
The derived parameter space indicates that a large abundance of PBHs with a mass around $1\ M_{\odot}$ and a fraction of PBHs to the total dark matter of $f_{\rm PBH} \gtrsim 0.2$ cannot explain the observed number of Icarus-like events and therefore is excluded. 
The methodology developed in this paper can be used to place tighter constraints on the fraction of PBHs from ongoing and future observations of ultrahigh magnification events.
\end{abstract}


\maketitle

\section{Introduction} \label{sec:intro}
Recent observations of highly magnified individual stars located near macro-critical curves of galaxy clusters open a new window to study small-scale structures \cite{2018NatAs...2..334K, 2022Natur.603..815W, 2023A&A...679A..31D, 2024arXiv240408045F}.
Icarus is the first example of such events observed in the MACS J1149 cluster with a magnification factor estimated as more than a thousand \cite{2018NatAs...2..334K}.
The peak magnification continues for less than two weeks, suggesting that the lensing magnification is primarily due to microlenses in the cluster, such as stars contributing to the intracluster light (ICL).
Icarus was observed during two-year observations with the Hubble Space Telescope (HST) and is the only event with such ultrahigh magnification detected throughout the observation period.

Because highly magnified images are observed near the critical curves, whose shapes and distributions reflect the mass distribution of lens objects, they have been subject to numerous theoretical studies.
The shape of a macro-critical curve, which arises from the overall density distribution of the main lens object such as a galaxy cluster, is perturbed by the existence of subhalos \cite{2024PhRvD.109h3517A, 2018ApJ...867...24D, 2024ApJ...961..200W}.
In addition to the macro-critical curve, micro-critical curves are formed around the location of microlenses such as ICL stars.
Both numerical simulations and analytic studies are conducted to investigate the effect of the microlenses on observational properties such as peak magnifications and high magnification event rates \cite{1986ApJ...306....2K, 2014ApJS..211...16V, 2017ApJ...850...49V, 2018ApJ...857...25D, 2018PhRvD..97b3518O, 2020AJ....159...49D, 2021arXiv210412009D, 2024A&A...687A..81P, 2024arXiv240316989V, 2024arXiv240408094W, 2024PhRvD.110h3514K}.
Among them, \citet{2024PhRvD.110h3514K} construct an analytic model for the probability distribution function (PDF) of the ultrahigh magnification tail.
Using their model, they calculate the expected number of Icarus-like events with magnifications exceeding a few thousand, showing good agreement with the results of HST observations.
Confronting these analytical studies with detailed observations provides a valuable approach to studying small-scale structures.

In addition to ICL stars, other compact objects such as primordial black holes (PBHs) can also contribute as microlenses.
PBHs are one of the possible alternatives to the standard cold dark matter (CDM) model (e.g., \cite{2020ARNPS..70..355C, 2021RPPh...84k6902C} for a recent review).
PBHs are thought to have formed in the early Universe, with one of the most widely studied mechanisms being the gravitational collapse of overdense regions during the radiation-dominant era following inflation \cite{1974MNRAS.168..399C, 1975ApJ...201....1C}.
In this scenario, one of the interesting characteristics is that PBHs can form across a wide mass range. 
While PBHs with masses smaller than $10^{15}\ {\rm g}$ would have evaporated by now due to Hawking radiation \cite{2010PhRvD..81j4019C}, those with masses larger than $10^{15}\ {\rm g}$ can survive to the present day and potentially contribute to the dark matter content of the Universe, which is our main focus on this study.
The mass of PBHs as well as the fraction of dark matter in PBHs are constrained by a wide range of targets, such as gravitational lensing \cite{2019NatAs...3..524N, 2020PhRvD.101f3005S, 2018PhRvD..97b3518O, 2007A&A...469..387T, 2001ApJ...550L.169A, 2019PhRvD..99h3503N, 2024Natur.632..749M, 2013PhRvL.111r1302G, 2014ApJ...786..158G, 2018PhRvL.121n1101Z, 2001PhRvL..86..584W}, dynamical processes \cite{2017PhRvL.119d1102K, 2016ApJ...824L..31B, 2020A&A...635A.107Z, 2014ApJ...790..159M, 2018MNRAS.478.3756C}, accretion \cite{2017PhRvD..95d3534A}, and gravitational waves \cite{2020PhRvD.101d3015V}.
While these studies suggest that PBHs may not account for the entire dark matter component, further probe of the parameter space is needed to better understand their possible contribution.

In the present paper, we calculate the expected number of Icarus-like events by using the analytic model for the high-magnification tail of the PDF \cite{2024PhRvD.110h3514K}.
Here we consider the relation between the radius and luminosity of source stars as well as the threshold for the apparent magnitude, which are not considered in \citet{2024PhRvD.110h3514K}.
Based on this realistic estimation, we find that ICL stars are consistent with our constraints on microlenses derived from the number of events observed in the HST observation.
Additionally, we constrain the mass and the mass fraction of PBHs under the assumption that PBHs follow the monochromatic mass function and that ICL stars exist.

In Sec.~\ref{sec:high_mag_events}, we first review the analytic model for the high-magnification tail of the PDF \cite{2024PhRvD.110h3514K} and the basic properties of the Icarus system studied in \citet{2018PhRvD..97b3518O}. 
Following this, we present the number of events and constraints on the parameter space of microlenses.
We then constrain the parameter space of the primordial black holes in Sec.~\ref{sec:pbh_constraint}.
We finally show the summary and discussions in Sec.~\ref{sec:summary_discussion}.

\section{Expected Number of highly magnified events} \label{sec:high_mag_events}
In this section, we show the expected number of Icarus-like events taking account of the source size as well as the threshold in apparent magnitude with the help of the recently developed analytic model for the high-magnification tail of the PDF presented in \citet{2024PhRvD.110h3514K}.
We briefly review the analytic model in Sec.~\ref{subsec:ana_model}, and the properties of MACS J1149 cluster and Icarus, which are mainly studied in \citet{2018PhRvD..97b3518O}, in Sec.~\ref{subsec:Icarus_sys}.
In Sec.~\ref{subsec:event_rate}, we show the estimation of the number of events as well as the constraint on the parameter space of microlenses.
Note that while \citet{2024PhRvD.110h3514K} calculate the number of events whose magnification exceeds a few thousands, they ignore the correlation between the source size and the intrinsic luminosity of a source star, which is considered in this paper.

\subsection{High-magnification tail of the PDF} \label{subsec:ana_model}
We consider a lens plane, where the total convergence and shear are represented by $\kappa_{\rm tot}$ and $\gamma_{\rm tot}$,  respectively, and satisfy the condition $\gamma_{\rm tot} = \kappa_{\rm tot}$.
This condition is applicable when we consider the matter distribution within a galaxy cluster, which can be approximated by a spherical isothermal density profile.
A mass fraction of microlenses is denoted by $f_{\star}$, leading to an average convergence due to microlenses of $\kappa_{\star} = f_{\star} \kappa_{\rm tot}$.
Under this scenario, the average magnification can be expressed as
\begin{equation}
    \mu_{\rm av} = \frac{1}{(1-\kappa_{\rm tot})^{2}-\gamma_{\rm tot}^{2}} = \frac{1}{1-2\kappa_{\rm tot}}.
\end{equation}
Here, the dependence of the mass fraction of microlens on the average magnification can be ignored \cite{2021arXiv210412009D}.

The high-magnification tail of the PDF, given the average magnification $\mu_{\rm av}$ and the mass fraction of microlenses $f_{\star}$, can be expressed as
\begin{eqnarray}
    &&\left.\frac{dP}{d\log_{10} \mu}\right|_{\rm high} = 2\ln 10\ P^{\rm PS}(\mu>\mu_{\rm th}) \nonumber \\
    && \hspace{20mm} \times \frac{1+e^{-1}}{1+\exp\left(\frac{\mu-\mu_{\rm max}}{\mu_{\rm max}}\right)}\left(\frac{\mu}{\mu_{\rm th}}\right)^{-2}.
\end{eqnarray}
Here, all microlenses are assumed to be uniformly and randomly distributed on the lens plane, each with the same mass $M_{\star}$, and consequently, identical Einstein radius
\begin{equation}
    \theta_{\rm Ein} = \sqrt{\frac{4GM_{\star}}{c^{2}}\frac{D_{\rm ls}}{D_{\rm os}D_{\rm ol}}}, \label{einstein_radius}
\end{equation}
where $G$ is the Newton constant, $D_{\rm ls}, D_{\rm os}$, and $D_{\rm ol}$ are the angular diameter distances between the lens and the source planes, the observer and the source plane, and the observer and the lens plane, respectively.
The integrated PDF for a point source above the magnification threshold $\mu_{\rm th}$ is denoted by $P^{\rm PS}(\mu>\mu_{\rm th})$ and is expressed as
\begin{eqnarray}
    && P^{\rm PS}(\mu>\mu_{\rm th}) \nonumber \\
    && \hspace{2mm} = \frac{A_{0}}{2.4} f_{\star} \kappa_{\rm tot} \sqrt{\mu_{\rm av}} \exp(-B_{0} f_{\star} \kappa_{\rm tot} \mu_{\rm av}) \left(\frac{10\mu_{\rm av}}{\mu_{\rm th}}\right)^{2}, \label{P_ps_A0_B0}
\end{eqnarray}
with the fitting parameters $A_{0} = 0.058$ and $B_{0} = 0.402$.
Note that the parameter combination $X = f_{\star} \kappa_{\rm tot} \mu_{\rm av}$ dictates the behavior of the integrated probability $Y = \sqrt{\mu_{\rm av}} P^{\rm PS}(\mu>\mu_{\rm th})$.
When $X \lesssim 1$, $Y$ scales linearly with $X$, defining what we refer to as the linear regime.
In contrast, in the nonlinear regime where $X \gtrsim 1$, the probability is exponentially suppressed. 
This suppression occurs due to the merging of micro-critical curves, which leads to an exponential decrease in the effective number of independent microlenses.
The maximum magnification, denoted by $\mu_{\rm max}$, depends on the source size $\sigma_{\rm W}$ as
\begin{equation}
    \mu_{\rm max} \simeq \mu_{\rm av} \sqrt{\frac{\theta_{\rm Ein}}{\sqrt{\mu_{\rm av}} \sigma_{\rm W}}} \min\left(1,\ (C_{0} f_{\star}\mu_{\rm av}\kappa_{\rm tot})^{-\frac{3}{4}} \right), \label{rmax_lin_nonlinear}
\end{equation}
with the fitting parameter $C_{0}=2.0$.

\subsection{MACS J1149 cluster and Icarus} \label{subsec:Icarus_sys}
\subsubsection{Average magnification} 
Icarus is a highly magnified individual star at redshift $z=1.49$ observed in the MACS J1149 galaxy cluster located at redshift $z=0.544$.
The average magnification $\mu_{\rm av}$ in the vicinity of Icarus can be expressed as a function of the distance $\beta$ from the macro-caustic of the galaxy cluster as
\begin{eqnarray}
    \mu_{\rm av}(\beta) =&& \mu_{\rm h} \mu_{\rm r} \frac{1}{\sqrt{2\pi(\sigma_{\rm ml}^{2} + \sigma_{\rm W}^{2})}} \nonumber \\
    &&\times \int_{0}^{\infty} d\beta' \sqrt{\frac{\beta_{0}}{\beta'}} \exp\left[-\frac{(\beta'-\beta)^{2}}{2(\sigma_{\rm ml}^{2} + \sigma_{\rm W}^{2})}\right], \label{Icarus_muav}
\end{eqnarray}
with $\mu_{\rm h} = 13$, $\mu_{\rm r} = 3$, and $\beta_{0}=0.045\ {\rm arcsec}$ \cite{2018PhRvD..97b3518O}.
The tangential component of the magnification can be expressed as a function of the distance $\theta$ to the macro-critical curve as
\begin{equation}
    \mu_{\rm t}(\theta) = \mu_{\rm h} \sqrt{\frac{\beta_{0}}{\beta(\theta)}} =\mu_{\rm h} \left(\frac{\theta}{\rm arcsec}\right)^{-1}. \label{Icarus_mut}
\end{equation}
The source size $\sigma_{\rm W}$ in units of arcsec is related to the physical source size $R_{\rm W}$ by
\begin{equation}
    \sigma_{\rm W} = 2.7 \times 10^{-12} \left(\frac{R_{\rm W}}{R_{\odot}}\right)\ {\rm arcsec}. \label{source_size}
\end{equation}
The effective source size, denoted by $\sigma_{\rm ml}$, which represents the variance of the random deflection angle by microlenses, is \cite{2021arXiv210412009D}
\begin{equation}
    \sigma_{\rm ml}^{2}(R_{\star},l_{\star}) = \kappa_{\star} \theta_{\rm Ein}^{2} \left(1-\gamma_{\rm E} + \ln \frac{2R_{\star}}{\theta_{\rm Ein}^{2}l_{\star}}\right),
\end{equation}
where $\gamma_{\rm E}$ is the Euler–Mascheroni constant.
The Einstein radius of a microlens with mass $M_{\star}$ is  
\begin{equation}
    \theta_{\rm Ein} = 1.8 \times 10^{-6} \left(\frac{M_{\star}}{M_{\odot}}\right)^{\frac{1}{2}}\ {\rm arcsec}. \label{theta_Ein}
\end{equation}
While $R_{\star}$ is originally given by $R_{\star} = \mu_{\rm av} \sigma_{\rm eff}$ with $\sigma_{\rm eff} = \sqrt{\sigma_{\rm W}^{2} + \kappa_{\star}\theta_{\rm Ein}^{2}}$, we set $R_{\star} = 1500$ following \citet{2021arXiv210412009D}.
We calculate $l_{\star}$ by using the relation $l_{\star} = 1/\sigma_{\rm eff}$.

The Icarus is observed at $0.13\ {\rm arcsec}$ away from the macro-critical curve.
The average magnification at this location is approximately $\mu_{\rm av} = 300$, which is significantly below the estimated magnification of $\mathcal{O}(10^{3})$.
This discrepancy suggests that the observed star is highly magnified due to the microlensing effect.
The ICL stars are one of the main contributions to the microlenses. 
The average convergence of the ICL stars is estimated as $\kappa_{\rm ICL} = 0.0046-0.0079$, depending on the assumed stellar initial mass functions \cite{2018NatAs...2..334K}.
Since the total convergence is estimated as $\kappa_{\rm tot} = 0.83$, the mass fraction of the ICL stars is calculated as $\kappa_{\rm ICL}/\kappa_{\rm tot} \simeq 0.005-0.009$.
The most probable mass of the ICL stars is $M_{\star} = 0.3\ M_{\odot}$ \cite{2018PhRvD..97b3518O}.

\subsubsection{Constraint on the radius of Icarus}
The radius of Icarus can be constrained by the peak magnitude and the source crossing time. 
For this purpose, we first need to specify the relation between the source size and its luminosity.
Assuming the blackbody with temperature $T$, this relation can be expressed using the Stefan–Boltzmann law 
\begin{equation}
    \frac{L}{L_{\odot}} = \left(\frac{R}{R_{\odot}}\right)^{2} \left(\frac{T}{T_{\odot}}\right)^{4}.\label{blackbody}
\end{equation}
The temperature is estimated as $T \simeq 12000\ {\rm K}$ from the spectral energy distribution, which corresponds to the blue supergiant \cite{2018NatAs...2..334K}.
The absolute magnitude can be calculated as
\begin{equation}
    M_{L_{\star}} \simeq 1.0 - 5\log_{10}\left(\frac{R}{R_{\odot}}\right),
\end{equation}
where we use the V-band magnitude of the sun, $M_{L_{\odot},V} = 4.83$, and the temperature of the Sun $T_{\odot} = 5777\ {\rm K}$, and include the bolometric correction, ${\rm B.C.} = -0.69$ \cite{1968ApJ...151..611M, 2013ApJS..208....9P}.
When the source is magnified with a factor of $\mu$, the apparent magnitude can be expressed as 
\begin{eqnarray}
    m_{L_{\star}} &=& M_{L_{\star}} + D - 2.5\log_{10}\mu \nonumber \\
    &\simeq& 45.1 - 5\log_{10}\left(\frac{R}{R_{\odot}}\right) - 2.5\log_{10}\mu, \label{mag_mu}
\end{eqnarray}
where $D = 5\log_{10}d - 5$ is the distance modulus with $d$ being the luminosity distance in units of parsec.
Considering the source redshift, the distance modulus is $D=45.2$.
Observations in the F125W band require the inclusion of the cross-filter K-correction to properly account for the differences in filter response. 
In the second equality, this correction is incorporated to ensure consistency in the analysis.
The observation of Icarus indicates that the peak apparent magnitude is $m_{L_{\star}} \lesssim 26$.
The threshold of the magnification $\mu_{\rm obsth}$ to achieve this observed apparent magnitude can be obtained from Eq.~\eqref{mag_mu} as
\begin{equation}
    \mu_{\rm obsth}(R) = 4.4 \times 10^{7} \left(\frac{R}{R_{\odot}}\right)^{-2}.
\end{equation}
We can see that smaller sources need to be magnified more.
Since the maximum magnification due to the micro-critical curves is obtained by Eq.~\eqref{rmax_lin_nonlinear} with Eqs.~\eqref{theta_Ein} and \eqref{source_size}, the condition $\mu_{\rm obsth}(R) \gtrsim \mu_{\rm max}$ yields the lower bound of the radius of the Icarus 
\begin{equation}
    R_{\rm min} = 69\left(\frac{\mu_{\rm av}}{300}\right)^{-\frac{1}{2}} \left(\frac{M_{\star}}{M_{\odot}}\right)^{-\frac{1}{6}}\ R_{\odot} \label{Rmin}.
\end{equation}
Note that we ignore the nonlinear suppression term in Eq.~\eqref{rmax_lin_nonlinear} to obtain Eq.~\eqref{Rmin}.
The source whose radius exceeds the above minimum can be observed.

The other constraint on the radius of Icarus can be obtained by the source crossing time, which is estimated by \cite{2018PhRvD..97b3518O}
\begin{equation}
    t_{\rm src} = 0.038\left(\frac{R}{R_{\odot}}\right) \left(\frac{v}{500\ {\rm km/s}}\right)^{-1}\ {\rm days}.
\end{equation}
There are three main contributions to the velocity $v$: the transverse peculiar velocity of a galaxy cluster, the relative transverse motion of a microlens within the galaxy cluster, and the transverse velocity in the source plane.
Among them, the transverse peculiar velocity of a galaxy cluster is the primary contributor \cite{2018PhRvD..97b3518O}.
The three-dimensional peculiar velocity of the MACS J1149 cluster is determined to be $v_{\rm 3D} = 638^{+1072}_{-393}\ {\rm km/s}$ with a 68\% confidence interval (CI) \cite{2016ApJ...831..110G}.
This is faster than the typical peculiar velocity since the MACS J1149 cluster is a merging cluster.
The angle between the merger axis and the plane of the sky is $\alpha \simeq 30\tcdegree$ \cite{2016ApJ...831..110G}, from which we can estimate the transverse peculiar velocity as $v_{\rm pec} = v_{\rm 3D} \cos\alpha$.
We assume that the probability distribution function of the (transverse) peculiar velocity follows the log-normal distribution to marginalize the uncertainty of the peculiar velocity in Sec.~\ref{sec:pbh_constraint}.
The observation of Icarus shows that the crossing time is less than 10 days, from which an upper limit on the radius can be obtained with the maximum radius being
\begin{equation}
    R_{\rm max} \simeq 260 \left(\frac{v_{\rm pec}}{500\ {\rm km/s}}\right) R_{\odot}. \label{Rmax_vpec}
\end{equation}
When the size of Icarus is larger than the above maximum radius, the duration of the peak magnification would continue larger, which is inconsistent with the observational results.

\subsubsection{Star population in the arc}
The width of the arc along the macro-critical curve, in which Icarus is located, is $w_{\rm arc} = 0.2\ {\rm arcsec}$.
The apparent magnitude of the arc is $25\ {\rm mag arcsec}^{-2}$ in the F125W band, which corresponds to the luminosity density $6.5\times 10^{9}\ L_{\odot}{\rm arcsec}^{-2}$.
We can convert the luminosity density into the number density of source stars that are subject to magnification.
Assuming a power-law luminosity function of stars, $dn_{\rm source}/dL \propto L^{-2}$, we can obtain the number density of source stars from the luminosity density as
\begin{equation}
    \frac{dn_{\rm source}}{d(L/L_{\odot})} = \frac{6.5\times 10^{9}\ {\rm arcsec}^{-2}}{\mu_{\rm av}\ln(L_{\rm max}/L_{\rm min})} \left(\frac{L}{L_{\odot}}\right)^{-2}.
\end{equation}
Here $L_{\rm min}$ and $L_{\rm max}$ are the minimum and maximum of the luminosity, respectively.
Using Eq.~\eqref{blackbody} with $T=12000\ {\rm K}$, we can rewrite the number density of source stars as a function of radius as 
\begin{equation}
    \frac{dn_{\rm source}}{d(R/R_{\odot})} = \frac{2n_{0}}{\mu_{\rm av}} \left(\frac{R}{R_{\odot}}\right)^{-3},
\end{equation}
with $n_{0} = 1.9\times 10^{7}\ {\rm arcsec}^{-2}$ assuming $L_{\rm min} = 0.1\ L_{\odot}$ and $L_{\rm max} = 10^{7}\ L_{\odot}$.
The maximum luminosity can be converted into the maximum radius of the source, $R_{\rm max} = 730\ R_{\odot}$.
We use these properties in the following section, where we calculate the expected number of the observed high-magnification events.

\subsection{Analytic estimate of event number} \label{subsec:event_rate}
Following the analytic model shown in Sec.~\ref{subsec:ana_model} and the properties of the Icarus system shown in Sec.~\ref{subsec:Icarus_sys}, we calculate the expected number of the highly magnified individual stars and predict their observed location.
The expected number of events in a single snapshot can be calculated by
\begin{eqnarray}
    N &=& 2\int^{\theta_{\rm max}}_{\theta_{\rm min}} d\theta\ \frac{2n_{0} w_{\rm arc}}{\mu_{\rm av}(\theta)} \int_{R_{\rm min}}^{{R_{\rm max}}} dR\ \frac{R_{\odot}^{2}}{R^{3}} \nonumber \\
    && \hspace{5mm} \times \int_{\mu_{\rm obsth(R)}}^{\infty} \left.\frac{dP}{d\log_{10} \mu}\right|_{\rm high}  d\log_{10} \mu, \label{Nevent}
\end{eqnarray}
where the coefficient 2 indicates that images appear on both sides of critical curves. 
We set the observation region from $\theta_{\rm min} = 0.0\ {\rm arcsec}$ to $\theta_{\rm max} = 1.3\ {\rm arcsec}$.
The maximum radius of sources is determined by the smaller value between Eq.~\eqref{Rmax_vpec} and $730\ R_{\odot}$.
The number of events described in Eq.~\eqref{Nevent} depends on three parameters: the peculiar velocity (or the maximum source radius), the mass fraction of microlenses, and the microlens mass.
Note again that we consider the same mass of the microlenses.
Although the average magnification in Eq.~\eqref{Icarus_muav} should, in principle, slightly depend on the mass fraction of microlenses and the microlens mass, we use the numerical result based on our fiducial parameters, $f_{\star} = 6 \times 10^{-3}$ and $M_{\star}=0.3\ M_{\odot}$, since we have confirmed that the parameter dependencies are relatively small.

\begin{figure*}
\includegraphics[width=\linewidth]{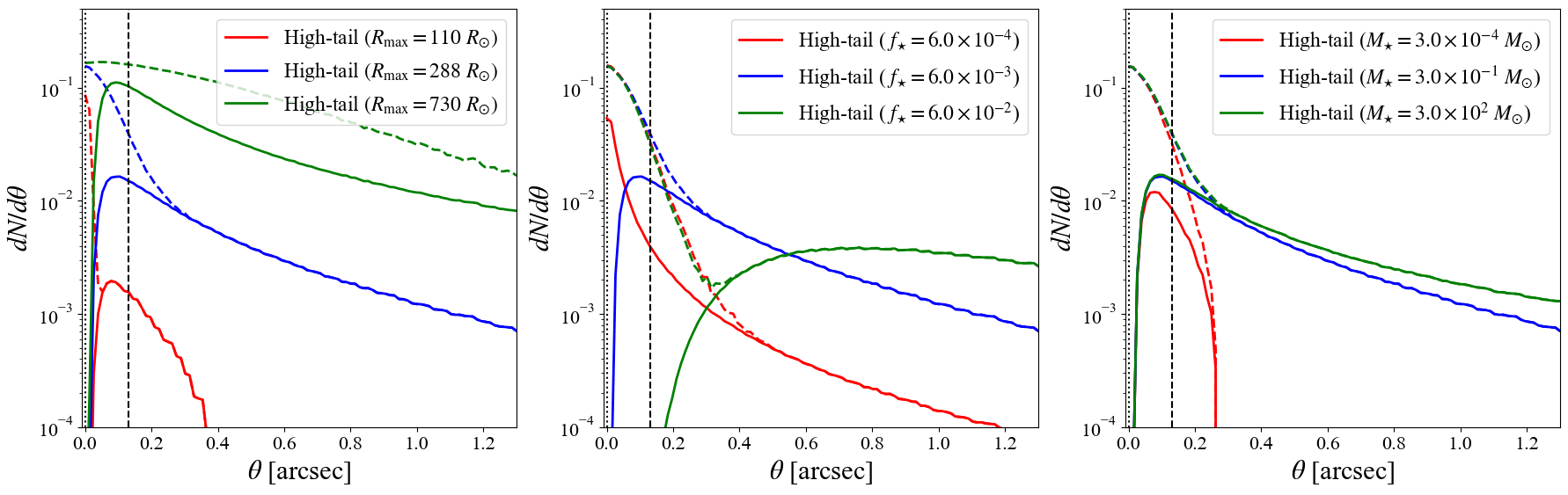}
    \caption{The expected number of the observed Icarus-like ultrahigh magnification events as a function of distance from the macro-critical curve. 
    The fiducial parameters used are $R_{\rm max}= 288\ R_{\odot}$, $f_{\star} = 6\times 10^{-3}$, and $M_{\star} = 0.3\ M_{\odot}$, represented by the blue lines. 
    Solid lines correspond to calculations using the high-magnification tail of the PDF, while dashed lines are based on the total PDF. 
    The vertical black dotted line indicates the position of the macro-critical curve, and the vertical black dashed line marks the location of Icarus. 
    Left: variation in the maximum source radius, reflecting the different peculiar velocities of the cluster. Middle: variation in the mass fraction of microlenses. Right: variation in the microlens mass.
    }
    \label{fig:dNdtheta}
\end{figure*}

Figure~\ref{fig:dNdtheta} shows the expected number of the observed highly magnified stars with different peculiar velocities of the MACS J1149 cluster, mass fractions of microlenses, and microlens masses.
The fiducial parameters are chosen to align with those of ICL stars \cite{2018PhRvD..97b3518O}.
Close to the macro-critical curve, the expected number of events increases due to the higher average magnification. 
However, at the center of the macro-critical curve, where the system enters the nonlinear regime, $f_{\star} \mu_{\rm av} \kappa_{\rm tot} \gtrsim 1$, the number of events is suppressed.
With the fiducial parameters, we find that the highest probability of observing highly magnified stars aligns with the position of Icarus.
A higher peculiar velocity leads to a larger maximum source radius, which, in turn, increases the expected number of events as can be seen in the left panel. 
As the mass fraction of microlenses increases, the turnover point shifts further from the macro-critical curve, as shown by the green solid line in the middle panel.
Conversely, when $f_{\star}$ is sufficiently small, the nonlinear regime disappears and there is no turnover, as demonstrated by the red solid line in the middle panel.
Since the minimum source radius $R_{\min}$ depends on the microlens mass, shown in Eq.~\eqref{Rmin}, a sufficiently small microlens mass reduces the integration range and results in a lower expected number of events shown in the right panel.
When the microlens mass becomes sufficiently larger, the expected number of events tends to converge.
This behavior is governed by the dependence of the microlens mass on the high-magnification tail of the PDF. 
Given that the number density of microlenses scales as $n_{\star}\propto M_{\star}^{-1}$ and the PDF scales as $dP/d\log_{10}\mu \propto n_{\star} \theta_{\rm Ein}^{2}$, the impact of the microlens mass on the high-magnification tail can be relatively small.

Note that the dashed lines represent results obtained using the total PDF, which includes the region around the average magnification modeled by a log-normal distribution, in addition to the high-magnification tail \cite{2024PhRvD.110h3514K}.
Near the center, where the average magnification is high enough to exceed the observational threshold, the contribution from the log-normal distribution dominates, leading to a higher expected number of events.
Given that the period of high magnification for Icarus is short, the event likely corresponds to one occurring in the high-magnification tail; otherwise, the duration of the peak magnification would be longer, as was the case for Earendel \cite{2022Natur.603..815W}. 
Therefore, in the following, we ignore the contribution of the log-normal distribution and only consider the high-magnification tail. 

By integrating the expected number shown in Fig.~\ref{fig:dNdtheta} with respect to the distance from the macro-critical curve $\theta$, we can obtain the total expected number of the observed high magnification events.
Considering that the Icarus is the single event with the peak magnification continuing for about two weeks during the HST two-year observation, we can estimate 52 independent snapshots, which we then multiply by our prediction from Eq.~\eqref{Nevent} for comparison with the observational results.

Given that the Icarus is the single event during the two-year observation, the observed mean number of the Icarus-like events, $\bar{N}$, can be constrained as follows. 
By assuming the Poisson distribution,
\begin{equation}
    P(N) = \frac{\bar{N}^{N}}{N!}e^{-\bar{N}},
\end{equation}
the probability that at least one event is observed can be expressed as $1-e^{-\bar{N}}$.
Therefore, the mean number of the events that is consistent with the observation is $\bar{N} \geq 0.051$ at the 95\% CI.

\begin{figure}
\includegraphics[width=\columnwidth]{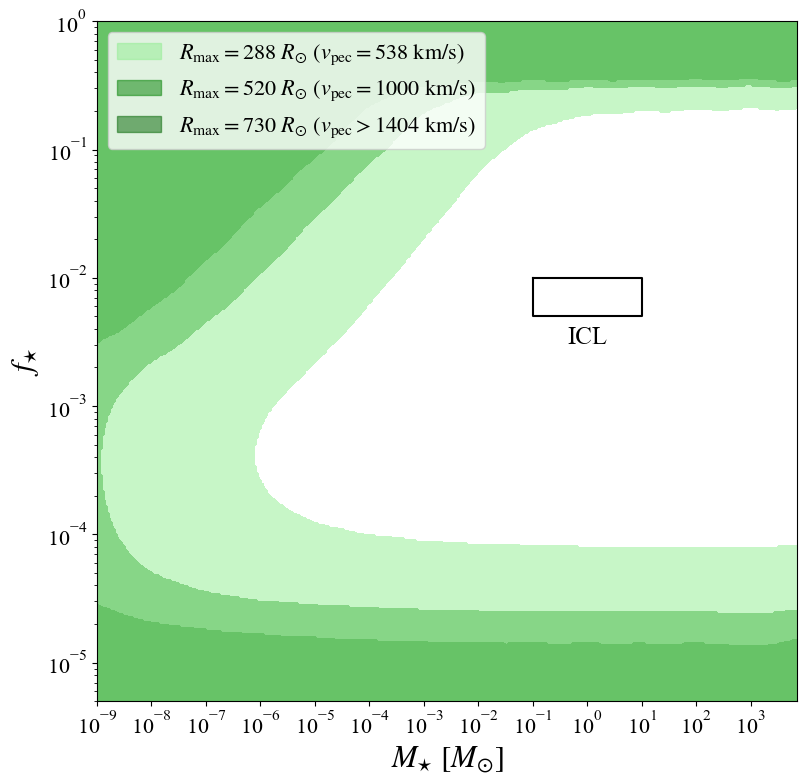}
    \caption{Constraints on the microlens parameters derived from the observed number of high-magnification events during the two-year HST observation period, assuming a single microlens component with a monochromatic mass function.
    The shaded regions indicate excluded parameter spaces at the 95\% CI.
    The three contours represent different source radii, which correspond to varying the peculiar velocities of the galaxy cluster.
    The parameter space of the ICL stars is shown in the black rectangular.
    }
    \label{fig:microlens_constraint}
\end{figure}

By comparing the expected number of events from our analytic estimation with the observed mean number of Icarus-like events, we can constrain the parameter space of the microlenses.
Figure~\ref{fig:microlens_constraint} shows the 95\% CI constraints on the mass and mass fraction of microlenses for different peculiar velocities of the galaxy cluster, assuming a single microlens component with a uniform mass.
As the peculiar velocity decreases, the number of expected events decreases, resulting in tighter constraints.
The results remain consistent with the properties of the ICL stars, indicating that the ICL stars alone can explain the observation of Icarus.
For a fixed microlens mass, the number of events is suppressed when the mass fraction of microlenses is either very small or very large.
The former suppression originates from the absence of microlenses in the linear regime, where the PDF is proportional to the number density of microlenses.
The latter suppression arises from the exponential damping of the PDF in the nonlinear regime.
For a fixed mass fraction of microlenses, the predicted number of events increases with the microlens mass and eventually plateaus at sufficiently high masses, leaving the upper limit of microlens mass unconstrained.

Note that \citet{2018PhRvD..97b3518O} also derive the microlens parameter space from the event rate (see Fig.~3 in their paper). 
There are two major differences between our approach and theirs.
First, the dependence of the microlens mass on the probability of high magnification differs significantly.
Our analytic model (shown in \citet{2024PhRvD.110h3514K}) shows almost no dependence on the microlens mass as mentioned in Sec.~\ref{subsec:event_rate}, whereas their model scales as $M_{\star}^{-1/2}$.
This results in lower microlens masses producing higher event rates in their framework.
Second, their approach overlooks the relationship between the source radius and the magnification threshold. 
As discussed earlier, smaller microlenses require a larger source radius to exceed the observed apparent magnitude, which results in a reduced number of events.
Due to these considerations, our constraints differ from theirs: while our constraints do not impose limits on larger microlens mass, their constraints fail to constrain the smaller microlens mass. 
Overall, we expect our constraints to be more precise.

\section{Constraint on PBH} \label{sec:pbh_constraint}
The PBHs are one of the possible candidates for the alternative to the standard cold dark matter.
The presence of the PBHs contributes to the microlens as well as the ICL stars.
In this section, we constrain the parameter space of the PBHs in the presence of ICL stars from the observed number of events with the marginalization over the uncertainty of the peculiar velocity of the galaxy cluster.
We denote $M_{\rm PBH}$ and $f_{\rm PBH}$ to describe the mass of PBHs and the mass fraction of PBHs relative to the total dark matter content, respectively.

\begin{figure}
\includegraphics[width=\columnwidth]{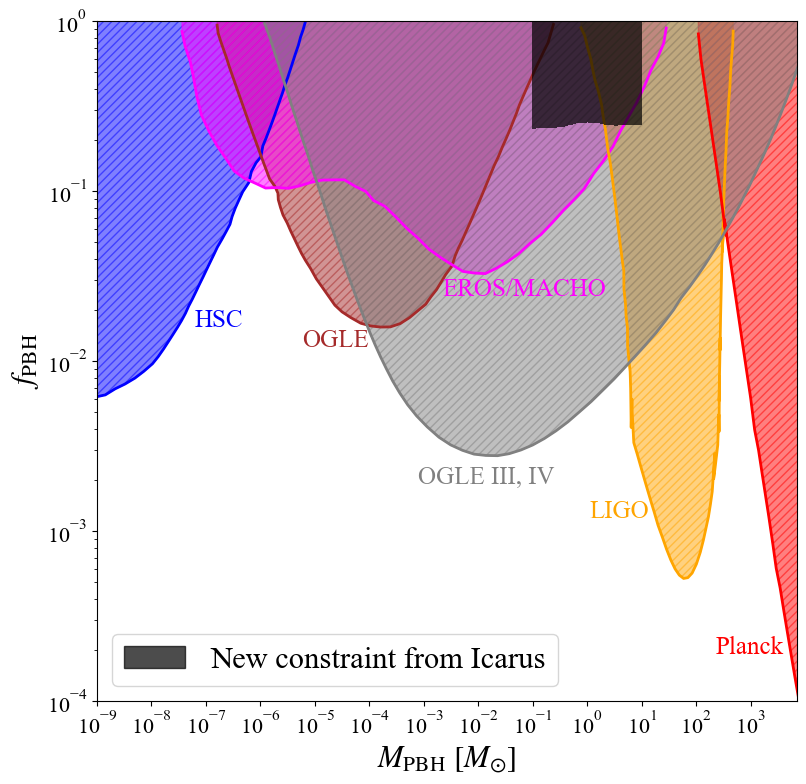}
    \caption{Constraints on the PBH parameters derived from the observed number of high-magnification events during the two-year HST observation period, shown in black shaded region. 
    The presence of ICL stars is taken into account in the constraints.
    The other constraints are obtained by HSC \cite{2019NatAs...3..524N} (blue), OGLE \cite{2019PhRvD..99h3503N} (brown), OGLE III, IV \cite{2024Natur.632..749M} (gray), LIGO \cite{2020PhRvD.101d3015V} (yellow), EROS/MACHO \cite{2007A&A...469..387T, 2001ApJ...550L.169A} (magenta), and Planck \cite{2017PhRvD..95d3534A} (red).
    All the constraints are shown at the 95\% CI.
    }
    \label{fig:PBH_constraint}
\end{figure}

Since the analytic model for the high-magnification tail of the PDF assumes that all microlenses have the same mass, our constraint is limited to the same mass range of ICL stars, $0.1 \lesssim M_{\rm PBH}/M_{\odot} \lesssim 10$.
Moreover, given the presence of ICL stars, the meaningful constraint comes only from the $f_{\rm PBH} \gtrsim f_{\rm ICL} \simeq 0.01$, because the suppression of the PDF in the nonlinear regime due to high $f_{\rm PBH}$ is not affected by the presence of ICL stars.
To marginalize the uncertainty of the peculiar velocity, we consider the log-normal distribution of the peculiar velocity as described in Sec.~\ref{subsec:Icarus_sys}.
In Fig.~\ref{fig:PBH_constraint}, we show the obtained constraint as well as the current constraints in the literature.
Our analysis shows that the mass fraction of PBHs is constrained to $f_{\rm PBH} \gtrsim 0.2$ for PBH masses ranging from $0.1\ M_{\odot}$ to $10\ M_{\odot}$.
Although our results are less stringent compared to previous studies, they remain consistent with them.

Note that \citet{2018PhRvD..97b3518O} also derive constraints on the parameter space of PBHs in the presence of ICL stars. 
Although their constraint uses the peak magnification rather than the number of events and therefore takes a different approach from ours, our results are largely consistent with theirs for microlens masses between $0.1\ M_{\odot}$ and $10\ M_{\odot}$. 
We also note that our new constraint takes account of the uncertainty of the peculiar velocity, while in \citet{2018PhRvD..97b3518O} the peculiar velocity is fixed to $v_{\rm pec}=500\ {\rm km/s}$.

\section{Summary and discussions} \label{sec:summary_discussion}
Using the analytic model for the high-magnification tail of the PDF developed in \citet{2024PhRvD.110h3514K}, we predict the expected number of highly magnified individual stars.
We focus on Icarus, a single event with peak magnification continuing less than two weeks near the macro-critical curve of the MACS J1149 cluster observed during the two-year HST observation.
To calculate the expected number of such events, we incorporate the relation between source radius and luminosity and take into account the constraints on the source radius derived from the peak magnitude, source crossing time, and total luminosity of the arc.
Under the assumption that all microlenses have the same mass, we find that the parameter space where the number of events aligns with observations is consistent with the presence of ICL stars. 
Additionally, we constrain the parameter space for PBHs to show that $f_{\rm PBH} \gtrsim 0.2$ for PBH masses between $0.1\ M_{\odot}$ and $10\ M_{\odot}$ is excluded at the 95\% CI.
This constraint is in agreement with previous studies.

Our constraint on the parameter space for microlenses and PBHs may be impacted by uncertainties related to the source temperature and the peculiar velocity distribution of the MACS J1149 cluster.
While the temperature of the Icarus is set to $T \simeq 12000\ {\rm K}$, there is an inherent uncertainty in the range of $11000\ {\rm K}$ to $14000\ {\rm K}$ \cite{2003IAUS..210P.A20C}.
This temperature variation affects the assumed blackbody luminosity, which in turn influences both the absolute and apparent magnitudes that determine the minimum source radius $R_{\rm min}$ and the magnification threshold $\mu_{\rm obsth}$.
A lower temperature results in larger $R_{\rm min}$ and $\mu_{\rm obsth}$, narrowing the integration range and reducing the expected number of highly magnified events.
Consequently, the parameter space excluded by the observation would expand.
Additionally, our assumption of a uniform temperature for other source stars may not be entirely accurate; adjusting this assumption could change the estimated number of events and thereby impact the constraints on the parameter space.
As shown in Fig.~\ref{fig:microlens_constraint}, variations in the peculiar velocity significantly shift the excluded parameter space. 
More precise measurements of the peculiar velocity, particularly if it aligns with the mean velocity, would tighten the constraints and potentially expand the excluded parameter space.

Although we only focus on the single Icarus event in this study, the observed number of highly magnified stars is rapidly increasing (e.g., \cite{2024arXiv240408045F}). 
We expect that tighter constraints will be obtained with similar analyses with more ultrahigh magnification events in the future.
In the Icarus system, for example, if the observed number of events is two (three), the excluded parameter space on the PBH would be $f_{\rm PBH} \gtrsim 0.08\ (0.01)$ at the 95\% CI.
As such, increasing the number of events would drastically shrink the allowed parameter space, highlighting the need for further exploration.

Although analyses of the number of highly magnified events prove to be an effective method for constraining the parameter space of microlenses and PBHs, the PBH constraint is limited to the same mass range as the ICL stars, as the analytic model for the high-magnification tail is based on a single microlens mass.
To broadly constrain the parameter space of the PBHs, it is essential to study the high magnification tail of the PDF with a bimodal mass function of microlenses.
We expect that these future studies will narrow the parameter space with a wide range of the microlens mass, leading to uncovering the stellar evolution history and the nature of dark matter.

\begin{acknowledgments}
This work was supported by JSPS KAKENHI Grant Numbers JP22J21440, JP22H01260, and JP20H05856.
H. K. thanks the hospitality of INAF OAS Bologna where part of this work was carried out.
\end{acknowledgments}

\bibliography{ref} 

\end{document}